\documentclass[prX,aps,tightenlines,preprint,floats,epsfig,superscriptaddress,nofootinbib]{revtex4}
\usepackage{epsfig}
\usepackage{amsfonts}
\usepackage{amsmath}
\usepackage{graphicx}
\usepackage{color}
\begin{document}
\title{Leptogenesis as a Common Origin for Matter and Dark Matter}
\author{Haipeng An}
\author{Shao-Long Chen}
\author{Rabindra N. Mohapatra}
\affiliation{Maryland Center for Fundamental Physics and Department of Physics, University of Maryland, College
Park, Maryland 20742, USA}
\author{Yue Zhang}
\affiliation{The Abdus Salam International Centre for Theoretical
Physics,  Strada Costiera 11, I-34014 Trieste, Italy}
\date{\today}
\begin{abstract}
We propose a model of asymmetric dark matter (DM) where the dark
sector is an identical copy of both forces and matter of the
standard model (SM) as in the mirror universe models discussed in
literature. In addition to being connected by gravity, the SM and
DM sectors are also connected at high temperature by a common set
of heavy right-handed Majorana neutrinos via their Yukawa
couplings to leptons and Higgs bosons. The lightest nucleon in the
dark (mirror) sector is a candidate for dark matter. The out of
equilibrium decay of right-handed neutrino produces equal lepton
asymmetry in both sectors via resonant leptogenesis which then get
converted to baryonic and dark baryonic matter. The dark baryon
asymmetry due to higher dark nucleon masses leads to higher dark
matter density compared to the familiar baryon density that is
observed. The standard model neutrinos in this case acquire masses
from the inverse seesaw mechanism. A kinetic mixing between the
$U(1)$ gauge fields of the two sectors is introduced to guarantee the
success of Big-Bang Nucleosynthesis.
\end{abstract}

\preprint{\vbox{\hbox{UMD-40762-471}}}
\preprint{\vbox{\hbox{UMD-PP-09-062}}}
\preprint{\vbox{\hbox{IC/2009/090}}}

\maketitle
\section{Introduction}

It now appears established that dark matter accounts for about one quarter
of the energy density, $\Omega$, of the universe and plays an essential role in
the formation of large scale structure in it.
The identity of dark matter, however, remains unknown since all the particles
in the successful standard model can be ruled out as candidates.
What the dark matter particles are, how
they interact with visible matter and how their relic abundance originates,
constitute some of the fundamental mysteries of
particle physics and cosmology today. Adding to this puzzle is the observation that
baryon contribution to $\Omega$ is only about a fifth (i.e., of the same order) of the dark matter contribution.
This raises the question: could the two have a common origin?

The most popular class of candidates for dark matter are the stable weakly interacting massive
particles (WIMPs), which arise in many well-motivated TeV scale
extensions of the standard model. Their stability is guaranteed by
some symmetry. While the WIMP dark matter does not decay due to the symmetry,
pairs of them can annihilate and their
relic density is determined by the freeze-out of the annihilation
from equilibrium to the SM particles. The fact that their observed
relic density can be naturally explained by the weak scale annihilation cross section
makes these models quite appealing. However in most models, the matter and dark matter
contributions to $\Omega$ are unrelated.

Coming to the question of the symmetry that ensures the stability of WIMPs, in most models one uses
a $Z_2$ symmetry (e.g., R-parity in supersymmetry or KK-parity in the case of
extra dimension models). On the other hand, one could quite easily imagine that
the stability of the WIMP is guaranteed by a continuous
$U(1)$ symmetry similar to baryon number (call it ``dark baryon number''). In such a case,
observed dark matter density would represent an asymmetry between dark matter and anti-dark matter
densities exactly as the case for the observed asymmetry between familiar matter and anti-matter.
If these two asymmetries could arise from a common mechanism, it would be a major step towards understanding
why their contributions to $\Omega$ are of the same order. It is the goal of this paper
to propose such a scenario.

The baryon-anti-baryon asymmetry in the universe (BAU) is of order
$10^{-10}$ and can arise from the laws of microphysics if the three conditions
proposed by Sakharov~\cite{Sakharov:1967dj} are satisfied, namely
baryon number violation, CP violation and out-of-equilibrium in the
early universe. The SM, however, fails to realize the third condition since it
requires that the Higgs mass must be less than 40 GeV; even if this condition
was satisfied, with three generations of fermions, it cannot explain why the
asymmetry is as large as the one observed. An elegant way to generate
the BAU is through leptogenesis~\cite{Fukugita:1986hr} in the
framework of seesaw mechanism~\cite{seesaw} which naturally explains
the smallness of the observed neutrino masses. In these models,
lepton asymmetry is generated through the out-of-equilibrium decay
of the very heavy right-handed (RH)  neutrinos which then get converted to baryon
asymmetry through the non-perturbative $B+L$ violating electroweak
sphaleron process~\cite{Kuzmin:1985mm}.

The appealing mechanism of baryogenesis via leptogenesis combined with the fact that the
relic abundances of baryons and dark matter are of the same order of
magnitude inspire us to think whether genesis of dark matter could also have its origin in
a manner similar to leptogenesis. The mirror universe models discussed
in the literature~\cite{mirror} appear to be a natural setting for this.
In the mirror models, the universe has two kinds of matter and forces:
one consisting of a standard model sector with forces and matter that we are familiar with,
such as quarks, leptons, W, Z, etc., and a parallel sector which
 is an exact replica (e.g., consisting of mirror duplicates of quarks and leptons, W, Z, $\gamma$, gluons and
the Higgs boson) called the mirror sector~\cite{mirror}. Forces (except for gravity) in one sector
do not affect the matter in the other sector. In what follows we will denote the
mirror particles by a prime on a symbol.
The two sectors communicate with each other by gravity and possibly some SM singlet
interactions which are very weak at the current age of the Universe.
There is a dark baryon and lepton number in the mirror sector which is the exact analog of the
familiar baryon and lepton number. The main hypothesis of this paper is
that the same leptogenesis mechanism that could be producing
matter-anti-matter asymmetry, is also producing asymmetry of dark matter-anti-dark-matter.
This then links the dark matter energy density of the Universe to that
contributed by matter making them of the same order, providing a resolution of the puzzle
stated in the beginning~\cite{adm}\footnote[1]{For earlier suggestion that mirror baryons constitute dark matter of the Universe, see~\cite{mdm}. Our model is, however, very different from these models in many respects}.

\begin{figure}[htb]
\begin{center}
\includegraphics[width=10cm]{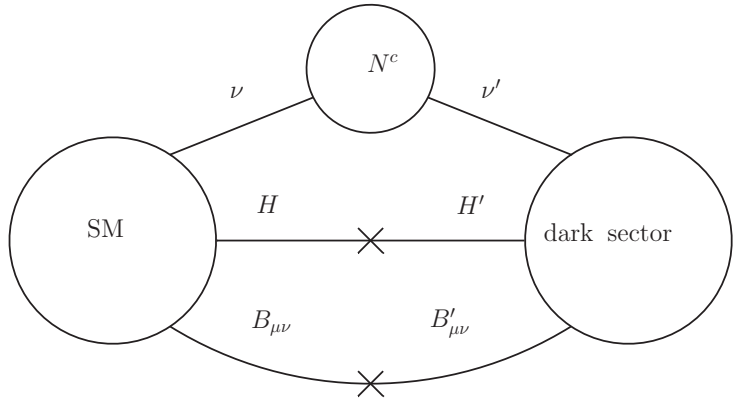}
\caption{The connection between the standard model and the dark
sector.}\label{po}
\end{center}
\end{figure}

A key ingredient in our attempt to connect the matter asymmetry to dark matter asymmetry
is the assumption that the visible and the mirror sectors talk to each other not only
through gravity but also through a common set of three right-handed
neutrinos coupled to leptons and Higgs fields in each sector through Yukawa couplings\cite{zurab}, as
shown in Fig.~\ref{po}. Since the RH neutrinos are standard model singlets, this is
consistent with gauge invariance. Also mirror symmetry makes the $N\ell H$ couplings on both sides equal.
The out-of-equilibrium decays of right-handed
neutrinos in the early universe can then produce lepton number asymmetries in
both sectors, which are then transferred to baryon and dark baryon
numbers through the sphaleron processes in each sector.
If one imposes exact mirror symmetry on the theory,
the primordial lepton asymmetries generated in each sector are equal and after sphaleron interaction
produce the same number density for baryons and dark baryons in the early universe. Since we expect the
symmetry breaking pattern in both sectors to be different for the model to be consistent with cosmology
(see below), the resulting energy density contributions can
be different and in the ratio $\Omega_B : \Omega_{DM} \approx 1:5$ if we require that mass of the dark
baryons is five times the mass of the familiar SM baryons. This
mass difference can arise from the difference in the scales of two
$SU(3)_c$ strong interactions ($\Lambda_{QCD},
\Lambda^\prime_{QCD}$). It turns out that this difference depends
on the ratio of two electroweak scales $v_{\rm wk}$
and $v'_{\rm wk}$, with $v_{\rm wk}:v'_{\rm wk}\sim 1:10^{3}$ giving the required difference between
$\Omega_{B}$ and $\Omega_{DM}$.

The spectra of the SM neutrinos $\nu_\alpha$ and the dark neutrinos
$\nu^\prime_{\alpha}$ are determined by the inverse
seesaw~\cite{inverse} and type-I~\cite{seesaw} seesaw mechanisms,
respectively, with the mixing between the $\nu_\alpha$ and
$\nu'_\alpha$ given by the ratio $v_{\rm wk}/v'_{\rm wk}$. The
dark neutrinos can decay into the SM particles due to this mixing.

In addition to the common set right-handed neutrinos, the SM
sector and the mirror sector can also be connected through Higgs
interaction and the kinetic mixing between the $U(1)$ gauge bosons
consistent with gauge invariance (as shown in Fig.~\ref{po}). The
photon sector mixing is necessary for the model to be consistent
with Big Bang Nucleosynthesis (BBN). In this work we assume the
photon in the mirror sector acquires a mass around 50 MeV through
the spontaneous symmetry breaking so that the mirror-electrically
charged particles which are heavier, pair annihilates into it
before the BBN, and the mirror photon itself decays to the
electron-positron pair through the kinetic mixing. To generate
this small mass two Higgs doublets are needed in the mirror
sector. The kinetic mixing between the photon and mirror photon
determines the lifetime of the mirror photon, so that success of
BBN puts a lower bound on this mixing. Furthermore, it
also determines the interactions between the dark matter
particles and the nucleons; therefore, using the current constraint
from the direct detection experiments produces an upper bound on
this mixing.

The common right-handed neutrinos generate a mixing between SM
neutrinos and mirror neutrinos. The exchange of mirror neutrinos
contribute to the neutrinoless double beta decay process ($0\nu\beta\beta$),
which also puts some mild constraints on the model.

\section{The framework}
The basic framework of the model, as already stated, is that there is a
mirror sector which before symmetry breaking is a complete duplication of
the standard model particles and forces. The form of the interactions as well
as the corresponding couplings in both sectors are identical. In this sense
prior to symmetry breaking, there are no new parameters in the matter
sector of the theory other than those in the standard model. The two sectors are
connected by a common set of three Majorana neutrinos which have Yukawa couplings to
both sectors as given below:
\begin{equation}
{\mathcal L}= -\lambda_{ij} \overline {N_j}P_L (\ell_i H)-\xi_{ij}
\overline {N_j}P_L (\ell^\prime_i H^\prime) -\frac{1}{2} \overline
{N_j}M_jN_j^c  + {\rm h.c.} \ ,
\label{yukawa}
\end{equation}
here $\ell$ and $H$ are the lepton and Higgs doublets in the SM,
$\ell^\prime$ and $H^\prime$ are their mirror counterparts; $\lambda$ and
$\xi$ are Yukawa couplings in both the two sectors. We assume mirror
symmetry so that all parameters in the mirror sector before symmetry
breaking are the same as those in the SM sector and, in particular, this would
mean that we have $\lambda =\xi\equiv h$.
In a subsequent section we will discuss other interactions connecting the
two sectors. Before discussing them, let us study the implications of
Eq.~(\ref{yukawa}) for genesis of familiar and dark baryon asymmetry of nature.
As noted, the latter will be identified with the dark matter.

\subsection{Genesis of baryonic and dark matter}
The canonical leptogenesis~\cite{Fukugita:1986hr} provides a
connection with the baryon asymmetry with neutrino masses via the
RH neutrinos needed for seesaw mechanism. In our model,
since we have a common set of RH neutrinos connecting to
leptons in both sectors, the leptogenesis connects the lepton
asymmetry in both sectors.

 Leptogenesis could be of two types. For the case of hierarchical right-handed neutrino
masses, $M_1\ll M_2, M_3$, a population of $N_1$ can be thermally
produced at temperature $T\sim M_1$ with negligible productions of
$N_{2, 3}$ and lepton asymmetry can be produced through
out-of-equilibrium decay of $N_1$ providing the interactions are
CP violating\cite{Buchmuller:2004nz}. On the other hand, we could
have at least two of the right-handed neutrinos highly degenerate,
in which case we have resonant leptogenesis\cite{Covi:1996wh}. In the first
case, the amount of lepton asymmetry is highly sensitive to the
values of the leptonic Yukawa couplings whereas in the latter
case, we could generate large lepton asymmetry regardless of the
coupling values by choosing the degree of degeneracy between the
RH neutrino masses. We will see below that within the constraints
of our model, the alternative of resonant leptogenesis provides a
more satisfactory framework. To see this, we first present the
formula for lepton asymmetry:
\begin{equation}
\epsilon\equiv  \frac{\sum_\alpha\left(\Gamma(N_1\to
H\ell_\alpha)-\Gamma(N_1\to \bar
H\bar\ell_\alpha)\right)}{\sum_\alpha\left(\Gamma(N_1\to
H\ell_\alpha)+\Gamma(N_1\to \bar H\bar\ell_\alpha)\right)}\,,
\end{equation}
where the sum goes only over the SM sector fields and applies to
both numerator and denominator. Correspondingly, $\epsilon^\prime$
in the mirror sector can be defined in the same way with $H, \ell$
replaced by $H^\prime, \ell^\prime$. The tree-level decay rates
are given by
\begin{equation}
\Gamma_{H\ell}\approx\Gamma_{\bar H\bar
\ell}\approx\frac{(\lambda^\dagger\lambda)_{11}M_1}{16\pi}\;,\;
\Gamma_{H^\prime\ell^\prime}\approx\Gamma_{\bar H^\prime\bar
\ell^\prime}\approx\frac{(\xi^\dagger\xi)_{11}M_1}{16\pi}\;,
\end{equation}
therefore the total rate is
$(\lambda^\dagger\lambda+\xi^\dagger\xi)_{11}M_1/(8\pi)$. The CP
asymmetric parameters $\epsilon^{(\prime)}$ originate from the
interference of tree-level and one-loop amplitudes. In our
scenario, the self energy diagram gets two contributions,
 due to the fact that both $\ell,
\ell^\prime$ show up in the loop. For hierarchical right-handed
neutrino spectrum, one obtains
\begin{equation}\label{e}
\epsilon\approx\sum_{k\neq 1}-\frac{1}{16\pi}\frac {M_1}{M_k}
\frac{{\mathcal Im}\{3[(\lambda^\dagger\lambda)_{k1}]^2+2[
(\lambda^\dagger\lambda)_{k1}(\xi^\dagger\xi)_{k1}]\}}
{(\lambda^\dagger\lambda)_{11}}\;,
\end{equation}
and $\epsilon^\prime=\epsilon(\lambda\leftrightarrow\xi)$. For the resonant leptogenesis case,
in the above expression, the factor $\frac{M_1}{M_2}$ is replaced by $\frac{M_1}{\Gamma_N}$
where $\Gamma$ is the decay width of the RH neutrino. The CP
asymmetries first generate asymmetries in the (dark) lepton
sectors. The evolution of this asymmetry can be studied using
Boltzmann equations~\cite{Buchmuller:2004nz}. In our case, they
differ from the conventional ones in the literatures since we
have included CP asymmetries due to the on-shell part of the
additional scattering processes. Other than this the discussion
is standard~\cite{Buchmuller:2004nz} and we do not reproduce it here.

Above the (mirror) electroweak scales, the (mirror) lepton numbers
will be transferred into a baryon asymmetry by the $SU(2)^{(\prime)}$
spharelon processes. The baryon asymmetry can be written as
\begin{equation}\label{et}
\eta_{b}\equiv\frac{n_B}{n_\gamma}\simeq \frac{3}{4}\cdot
\frac{g_{\ast s}^0}{g_{\ast s}}\cdot \frac{28}{79}\cdot
\kappa\cdot \epsilon\,,
\end{equation}
where $g_{\ast s}^0$ and $g_{\ast s}$ are the total number of
relativistic degrees of freedom today and during sphaleron,
${28}/{79}$ accounts for the sphaleron effects, $\kappa$ is the
washout factor.

The resulting ratio of baryon to dark baryon is
\begin{equation}
r \equiv \frac{n_B}{n_{B}'}=\frac{\kappa}{\kappa'} \cdot
\frac{\displaystyle\sum_{k\neq 1} \frac{1}{M_k} {\rm
Im}\{3[(\lambda^\dagger\lambda)_{k1}]^2+2[(\lambda^\dagger\lambda)_{k1}
(\xi^\dagger\xi)_{k1}]\}}{\displaystyle\sum_{k\neq 1}
\frac{1}{M_k} {\rm
Im}\{3[(\xi^\dagger\xi)_{k1}]^2+2[(\lambda^\dagger\lambda)_{k1}
(\xi^\dagger\xi)_{k1}]\}}\,,
\end{equation}
which means the baryon to dark matter relic abundance satisfies
\begin{equation}\label{ratio}
\frac{\Omega_{\rm baryon}}{\Omega_{\rm dark\ matter}} =
\frac{m_N}{m_{DM}} \cdot r\,,
\end{equation}
where $m_N, m_{DM}$ are the nucleon and dark baryon masses,
respectively. For the resonant leptogenesis case, the factor $\frac{1}{M_k}$
is replaced by $\frac{1}{\Gamma_N}$.

Mirror symmetry implies that $\epsilon=\epsilon'$ and
$\kappa=\kappa'$. This means $r=1$, the ratio of relic abundance
is totaly determined by the mass ratio of nucleon and dark matter
particle, i.e., $m_N/m_{DM}\sim 1/5$. We will see in the next
section how the mass differences between the nucleons in the two
sectors can arise dynamically.

There are several constraints on the
parameters of the model so that an adequate lepton number is
generated and the washout effects due to scattering and decays do
not reduce these values below what is required by observations. We
summarize them below (as noted, we have denoted the leptonic Yukawa
couplings by $h=\xi=\lambda$).

\begin{itemize}
\item We need to make sure that the resulting lepton asymmetries
do not get erased by scattering processes of the type $\ell +
H\leftrightarrow \bar{\ell}+\bar{H}$ via right-handed neutrino
exchange~\cite{zurab}. These processes go like $\frac{(h^\dagger
h)^2 T^3}{M_R^2}$ and for them to be out of equilibrium at $T\sim
M_R$, we require $M_R \geq \frac{(h^\dagger h)^2 M_{Pl}}{120
\pi}$;

\item The decays of the RH neutrinos contribute to the
washout factor $K \equiv \frac{\Gamma_N}{H(M_R)}
\sim \frac{(h^\dagger h)M_{Pl}}{12 \pi \sqrt{g_\ast}M_R}$ and should be smaller than
about $10^6$;

\item The amount of primordial lepton asymmetry for the hierarchical RH neutrino case
is roughly given by $\epsilon < \frac{h^\dagger h}{16\pi}$;

\item Finally, we borrow a result from a subsequent section about
mirror masses that in order to retain the success of BBN in our
model, the mirror neutrinos must decay before the epoch of
nucleosynthesis to ordinary leptons, which requires that we must
have $m_{\nu'} \geq 100$ MeV. This translates into a constraint on
the Yukawa couplings and masses of right-handed neutrinos as
follows: $\frac{h^2}{M_R} \geq \frac{0.1 {\rm GeV}}{v^{\prime 2}_{wk}}$.
\end{itemize}

The consequence of these constraints is that for the hierarchical
RH neutrino case, it is not possible to get enough primordial
lepton asymmetry while simultaneously satisfying the washout and
mirror neutrino mass constraints. We therefore adopt the resonant
leptogenesis with $M_1\simeq M_2 \sim 10^{8}$ GeV in which case
there is an enhancement factor of $M_R/\Gamma_N$, so that it
compensates for small Yukawa coupling effects.

\subsection{The complete model}
In this subsection, we describe the new features that go beyond the usual mirror models
(i.e., SM and its duplicate in each sector) that are needed for
consistency. In all the new features, we maintain the exact mirror symmetry for all dimension four terms. For simplicity
we may add soft mirror symmetry breaking terms, which may arise from a mirror symmetric model at high scale via
spontaneous symmetry breaking. The presence of the soft breaking terms allows us to have symmetry breaking patterns in
the two sectors different while the interactions remain symmetric.
The model presented is non-supersymmetric but its supersymmetric extension will preserve its main
features.

Instead of one Higgs doublet as in the standard model,
we consider two Higgs doublets $H_{u,d}$ and one $Y=2$ triplet $\Delta$ to both sectors.
And to avoid large flavor changing neutral current, a $Z_2$ symmetry is imposed in each
sector so that the up-type fermions only couple to $H_u$ or
$H_u'$ whereas the down-type fermions to $H_d$ or $H_d'$.

The triplet Higgs bosons couple to left-handed lepton pairs as
\begin{equation}
{\cal L}_{T} = - Y_\Delta \bar {\ell^c} \Delta \ell -
Y'_{\Delta} \bar {{\ell'}^c} \Delta' {\ell'} + {\rm h.c.}
\end{equation}
The neutral components of the triplet Higgs bosons get vacuum expectation values (VEVs)
after spontaneous breaking of electroweak symmetries.
Therefore, the left-handed neutrinos in the two sectors
obtain Majorana mass terms
and the generic neutrino mass matrix is given as in eq.~(\ref{inv}).
We will assume that the triplet VEV (or type II) contribution to
the mass of the known neutrinos is zero whereas it is non-zero for the
mirror neutrinos.

A few comments about the Higgs sector and the vev of Higgs fields
is in order. First, in order to get the mirror doublet vevs to be
larger than the familiar SM Higgs, we need to include soft
breaking mass terms e.g.
\begin{eqnarray}
{\cal L}_{soft}~=~\sum_a M^2_a H^\dagger_aH_a~+~\sum_a M^{'2}_a
H^{'\dagger}_aH^{'}_a~+ m^2_T
\vec{T}^{\dagger}\cdot\vec{T}~+~m^{\prime 2}_T
\vec{T}^{\prime\dagger}\cdot\vec{T^\prime}
\end{eqnarray}
where $M^2_a,~M^{'2}_a, ~m^{\prime 2}_T$ are alls negative and
$|M^{'2}_a| >> |M^2_a|$. As a result, the mirror Higgs vevs become
different from the familiar SM Higgs and its partner.

Note also that since the Higgs triplet of the visible sector has
no vev, it is free of constraints from electroweak radiative
correction constraints. Admittedly, such differences in vev
require fine tuning of parameters, whose proper understanding is
beyond the scope of this paper, where we discuss the new scenario
for dark matter and not the naturalness of the values of all the
parameters of the theory. In deriving our conclusions, we have
taken into account all the allowed renormalizable couplings
involving the Higgs fields. We don't display them here for
brevity since they do not affect our discussion in this work.

\subsection{The mirror nucleons}

Generally, the mass of the nucleon is composed by two parts, namely the
trace anomaly part and the quark masses. The trace anomaly part is
proportional to the hadronic scale $\Lambda_{QCD}$ and the
$\beta$-function of QCD, and the contribution from the quark masses
are proportional to the quark masses. Take the proton as an example; we
have
\begin{equation}
m_p = k \Lambda_{QCD} + 2 h_{pu} m_u + h_{pd} m_d\ ,
\end{equation}
where $h_{pu}$ and $h_{pd}$ can be determined from the pion-nucleon
$\sigma$-term. In the SM, since $m_u$ and $m_d$ are much smaller
than the hadronic scale, their contribution can be neglected. In the
mirror sector the contribution of the mirror quark masses may not be
negligible, so masses of the mirror proton and neutron can be written as
\begin{eqnarray}
m_{p'} &=& k_p'\Lambda_{QCD}' + 2 h_{pu}' m_{u'}+h_{pd}'
m_{d'}\,,
\nonumber\\
m_{n'} &=& k_n'\Lambda_{QCD}' + 2 h_{nd}' m_{d'}+h_{nu}'
m_{u'}\,,
\end{eqnarray}
where we set different coefficients of proportionality for the
mirror proton and neutron since if the masses of the mirror quarks
are comparable with the mirror hadronic scale, the isospin
symmetry is strongly broken and there is no reason to set them to
be the same. Under exact mirror symmetry, to get the correct dark
matter relic density one must have the mass relation: $m_{p'}
\approx 5 m_p$ for mirror proton as the dark matter particle, or
$m_{n'} \approx 5 m_p$ for dark neutron as dark matter. This
difference in the nucleon masses in the two sectors can arise if
the two electroweak scales are different with $v'_{wk}\gg v_{wk}$.
What this difference of EW scales does is to make the mirror
quarks ($t', b', c', ..$ etc.) much heavier than the familiar
quarks of SM. If we further assume the two strong interaction
coupling constants become equal at high scale due to mirror symmetry,
we can get a relation between the two electroweak scales and the
two hadronic scales using one loop evolution of the QCD couplings
in the two sectors and assuming $\Lambda'_{QCD}$ and
$\Lambda_{QCD}$ to be the scales where the QCD couplings
$\alpha_{QCD}$ become of order one. Under different conditions for
the mirror quark spectra, we get the following relations for the
ratio ${v_{\rm wk}}/{v_{\rm wk}'}$.
\begin{eqnarray}\label{ratio}
&&\left(\frac{v_{\rm wk}}{v_{\rm wk}'}\right)^{4} =
\frac{\Lambda_{QCD}^9(m_um_dm_s)^{2/3}}{\Lambda_{QCD}'^{11}}
\left(\frac{\sin\beta'}{\sin\beta}\right)^{4}
\left(\frac{\tan\beta'}{\tan\beta}\right)^{-2}, \ \ \
{\rm for\ \ }\Lambda_{QCD}'<m_{u'}, m_{d'}; \nonumber\\
&&\left(\frac{v_{\rm wk}}{v_{\rm wk}'}\right)^{5} =
\frac{\Lambda_{QCD}^{27/2}(m_dm_s)}{\Lambda_{QCD}'^{31/2}}
\left(\frac{\sin\beta'}{\sin\beta}\right)^{5}
\left(\frac{\tan\beta'}{\tan\beta}\right)^{-3}, \ \ \
{\rm for\ \ }m_{u'}<\Lambda_{QCD}'<m_{d'}; \nonumber \\
&&\left(\frac{v_{\rm wk}}{v_{\rm wk}'}\right)^{5} =
\frac{\Lambda_{QCD}^{27/2}(m_um_s)}{\Lambda_{QCD}'^{31/2}}
\left(\frac{\sin\beta'}{\sin\beta}\right)^{5}
\left(\frac{\tan\beta'}{\tan\beta}\right)^{-2}, \ \ \
{\rm for\ \ }m_{d'}<\Lambda_{QCD}'<m_{u'}; \nonumber \\
&&\left(\frac{v_{\rm wk}}{v_{\rm wk}'}\right)^{4} =
\frac{\Lambda_{QCD}^{27/2}m_s}{\Lambda_{QCD}'^{29/2}}
\left(\frac{\sin\beta'}{\sin\beta}\right)^{4}
\left(\frac{\tan\beta'}{\tan\beta}\right)^{-2},  \ \ \ {\rm for\ \
}m_{u'}, m_{d'}<\Lambda_{QCD}'<m_{s'}\ .
\end{eqnarray}
where $m_{u'}< m_{d'}$ if ${\tan\beta'}/{\tan\beta} < {m_d}/{m_u}$ and
$m_{u'}>m_{d'}$ for ${\tan\beta'}/{\tan\beta}
> {m_d}/{m_u}$. This result does not depend on whether $m_{c'}< m_{b'}$
or $m_{c'}>m_{b'}$. From Eq. (\ref{ratio}) one can see that
$\Lambda'_{QCD}$ grows slowly with the increasing of $v'_{wk}$,
that $\Lambda'/ \Lambda\approx (v'_{wk}/v_{wk})^{1/3}/10$, there
$v'$ increases a lot if $\Lambda'_{QCD}$ is a few times larger
than $\Lambda_{QCD}$ which means $u'$ and $d'$ in the mirror
nucleon might be nonrelativistic. In QCD, if the masses of $u$ and
$d$ quarks were much larger than the hadronic scale they would be
nonrelativistic inside the baryons. In that case the mass of the
nucleon would be approximately equal to the sum of the quark
masses plus a negative potential generated by the gluon field.
Therefore, we can conclude that as $v'_{wk}$ grows larger and
larger $h'_{u,d}$ approach $1$ and $k'$ gets smaller and
smaller. Using lattice QCD one can calculate $k'$ and $h'_{u,d}$
with different values of $v'_{wk}$. In this work, since we will
see that the masses of $u'$ and $d'$ are comparable with the
mirror hadronic scale, we assume that $h_{pu}'
=h_{pd}'=h_{nu}'=h_{nd}'=1$. For the other contribution, from the
above analysis we know that as $v'_{wk}$ goes up it will get
smaller, then crossover zero and then get negative, so in
this work we will neglect this contribution.

In the SM, the neutron is slightly heavier than the proton due to $m_u<m_d$
and a free neutron will decay to a proton through beta decay. With the
exact mirror Yukawa couplings, the dark neutron is expected to be
heavier than the dark proton as in the SM. The situation could be
different for two Higgs doublets $H^{(\prime)}_{u,d}$, with
$\tan\beta=v_u/v_d, \tan\beta'=v'_u/v'_d$ different. $H^{(\prime)}_u$
only couples to up-type (mirror) quarks and (mirror) neutrinos while
$H^{(\prime)}_d$ to down-type (mirror) quarks and charged (mirror)
leptons, respectively, by imposed $Z_2$ symmetries in both the two
sectors. As we can see below (Table I) , when
$\tan\beta'/\tan\beta>m_d/m_u$, the dark neutron is lighter than
the dark proton.

In Table I, $\Lambda_{QCD}'$, the $u'$, $d'$, $s'$ and nucleon
masses for different choices of $\tan\beta'$ are listed, while
demanding the mass of the lightest nucleon in the mirror sector to
be 5 GeV.  It turns out $\Lambda'_{QCD}$ depends on
$\tan\beta'$ very mildly, while the dark weak scale $v_{\rm wk}'$
increases with $\tan\beta$. For low $\tan\beta'$, $m_{u'}<m_{d'}$, the dark
proton is dark matter, and for larger $\tan\beta'\gtrsim2\tan\beta$,
the  dark neutron is dark matter. In the following calculation we
will take the case $\tan\beta=50$, $\tan\beta'=200$. Therefore,
the mirror neutron is the dark matter particle.

\begin{table}[htb!]
\centerline{
\begin{tabular}{|c|c|c|c|c|c|c|c|}
\hline $\tan\beta'$ & $v_{\rm wk}'$ (TeV) & $\Lambda'_{QCD}$(GeV) &
$m_{u'}$(GeV) & $m_{d'}$(GeV) & $m_{e'}$(GeV) & $m_{p'}$(GeV) & $m_{n'}$(GeV) \\
\hline
50 & 123 & 1.06 & 1.25 & 2.5 & 0.25 & 5.0 & 6.25\\
100 & 164 & 1.04 & 1.67 & 1.67 & 0.17 & 5.0 & 5.0\\
200 & 246 & 1.06 & 2.5 & 1.25 & 0.13 & 6.25 & 5.0\\
500 & 369 & 1.27 & 3.57 & 0.71 & 0.07 & 7.85 & 5.0\\
\hline\end{tabular}} \caption{The values of $v'_{\rm wk},
\Lambda'_{QCD}$ for different $\tan\beta'$ with the light mirror
nucleon mass fixed at 5 GeV. The other inputs are taken as
$\tan\beta=50$, $\Lambda_{QCD}=200$ MeV, $m_u=2.5$ MeV, $m_d=5$ MeV.}
\end{table}

\section{Neutrino masses}

The neutrino mass matrix in the basis $(\nu, \bar N^c, \nu^\prime)$
can be written as
\begin{equation}\label{inv}
{\mathcal M}= \left(\begin{array}{c c c}
   \mu  &   M_D & 0 \\
  M_D^T   & M_R   & M^{\prime T}_D\\
  0 & M^\prime_D     &  \mu'
\end{array}\right)\ ,
\end{equation}
where $M_D=\lambda v_{\rm wk}$ and $M_D'=\xi v'_{\rm wk}$, $\mu$
and $\mu'$ are Majorana mass matrices for SM neutrinos and mirror
neutrinos generated from type-II seesaw mechanism, respectively
We take the exact mirror symmetry for Yukawa couplings, i.e.,
$\lambda=\xi=h$. Since we want to roughly reproduce the general features of the
neutrino masses and the mixing of the SM neutrinos,
 the flavor indices are suppressed in the following
calculations.

From Table I $v'_{\rm wk}$ is chosen about $10^3$ times larger
than $v_{\rm wk}$, so $M_D'$ is $10^3$ times larger than $M_D$ due
to the mirror symmetry of the Yukawa couplings. If $\mu' M_R\ll
M'^2_D$, the SM neutrino mass lies in the type II + inverse seesaw
regime, $M_\nu\approx \mu + M_D M^{\prime-1}_D \mu' M^{\prime
T-1}_D M_D^T=\mu +\mu' (v_{\rm wk}/v'_{wk})^2$, while the mass of
the dark sector neutrino receives dominant contribution from type I
seesaw mechanism, i.e., $M_\nu' \approx\mu' - M^\prime_D M_R^{-1}
M^{\prime T}_D\approx - M^\prime_D M_R^{-1} M^{\prime T}_D$. On
the other hand, if $\mu' M_R>M'^2_D$, the SM neutrino is
contributed from type I and type II seesaw mass $M_\nu'\approx
\mu' - M_D M_R^{-1} M^{T}_D$, while the dark neutrino mass is
mainly type II, $M_\nu'\approx\mu'$. However, in the latter case,
the triplet Higgs contribution dominates the leptogenesis in the
dark sector~\cite{tripletlepto} which will ruin the relation
$n_B=n_{B'}$. Hence, in this work we adopt the first scenario.
Also, since $\mu$ is determined by the interaction between doublet
Higgs bosons and the triplet Higgs boson in the SM sector, it is
independent of other quantities in the neutrino mass matrix;
therefore, we can simply assume it to be much smaller than the
neutrino masses and neglect its contribution.

In order to generate the relic density of baryon and dark matter, and
to avoid the constraints from various experiments and
observations which will be discussed in the following sections, we
choose the following parameters:
\begin{eqnarray}\label{par}
M_R \approx 1 \times 10^{8} {\rm GeV}, \; h=0.015,\;
v'_{\rm wk}/v_{\rm wk} =10^3, \;\mu' \approx 100 {\rm KeV}\ .
\end{eqnarray}
The resulting neutrino masses are of order $M_\nu \approx 0.1$ eV,
$M_\nu'\approx 150$ MeV. We note that there is a non-trivial mixing
between the SM and dark neutrinos which is $\hat \nu = \nu +
U_{\nu\nu'} \nu'$, $U_{\nu\nu'}\approx M_D/ M_D'\approx 1.0 \times
10^{-3}$; i.e., the mixing between familiar and mirror neutrinos is
a universal number for all flavors in the flavor basis. The dark
neutrino, once produced, can decay to $e^+e^-\nu$ due to
this mixing through weak interactions. For
the parameters set in Eq.~(\ref{par}) the lifetime of dark
neutrinos can be estimated as $\tau_{\nu'} < 0.5$ sec.
Therefore the mirror neutrinos decay to the SM sector before
the BBN epoch.

We wish to emphasize that the set of parameters in Eq.~(\ref{par})
is quite unique. When we fix the ratio of $v_{\rm wk}/v_{\rm wk}'$
from Eq.~(\ref{ratio}), the SM neutrino masses and mixings are
completely determined by $\mu^\prime$. Furthermore, we cannot make $M_R$
heavier because that will reduce the mass for dark neutrino
thereby contradicting  with the constraints from BBN. On the other
hand, for $M_R= 10^{8}$ GeV, we have the $K$ factor relevant for
leptogenesis $K=M_D M_R^{-1}M_D^T/m_\ast \approx {1.3\cdot 10^{-7}
\rm{GeV}}/{10^{-3}\rm{ eV}} =1.3\times 10^5$, which implies a
washout factor $\kappa \approx 1\times 10^{-6}$. According to
Eqs.~(\ref{e}) and (\ref{et}), to get $\eta_b\approx5\cdot
10^{-10}$, we need to have a value of primordial CP asymmetry
$\epsilon\sim 0.05$. In this scenario, as already noted, leptogenesis can be
realized if  at least two right-handed neutrinos are
quasi-degenerate~\cite{Covi:1996wh}. Decreasing $M_R$ will
increase the $K$ factor; larger $K$ implies stronger washout
effect or smaller $\kappa$, threatening the success of
leptogenesis.

\section{The constraints}
\subsection{Constraints from BBN}
BBN is well constrained by the expansion
rate of the universe at the temperature $T_{BBN}\sim 1$ MeV. The
Hubble expansion rate is determined by the total energy density,
which constrains the light degrees. When at $T\sim 1$ MeV, we have
the number of degrees of freedom (d.o.f.) $g_\ast=10.75$ contributed
from the SM photons, electrons and neutrinos. The constraints on new
d.o.f. is conventionally quoted as $\Delta N_\nu$, the effective
number of additional light neutrino species. A reliable bound is
$\Delta N_\nu\leq 1.44$ at 95\% CL by various present
observations~\cite{Cyburt:2004yc}.

In the symmetric mirror model, the mirror neutrinos as well as the
morror photon and electrons will contribute another $10.75$ to
$g_\ast$, thereby completely spoiling the success of BBN. One way
to avoid the BBN bound is to have the hidden sector with lower
temperature than the SM sector. This could be achieved if the
reheating temperatures after inflation the two sectors are
different\cite{bere}. However, this scenario does not work here
since the SM sector and the mirror sector are
connected by the common right-handed neutrinos which can bring the
two sectors back into thermal equilibrium even if they have
different couplings with the inflaton. When the temperature is
lower than the mass of the lightest right-handed neutrino, the
interaction between the Higgs bosons in two sectors
$\lambda|H|^2|H'|^2$ and the kinetic mixing between the $U(1)_Y$
gauge bosons $\frac{\varepsilon_B}{2} B^{\mu\nu}B'_{\mu\nu}$  (see Appendix)
provide two alternative mechanisms to keep the two sectors in
thermal equilibrium. The kinetic mixing of the $U(1)_Y$ bosons of
the two sectors $B$ and $B'$ induces the kinetic mixing between
the photon and mirror photon; therefore if mirror photon is
massless there is a long range interaction between the two
sectors. As a result the two sectors will never be thermally
decoupled from each other thereby increasing the number of degrees
of freedom at BBN to unacceptable values.

In our model, since the mirror neutrinos have masses above a 100 MeV
 and lifetimes much less than a second, they do not pose any
problem for BBN. We only have to make sure that the photon
contribution is eliminated. In order to achieve this, we work with
two-Higgs-doublets in each sector such that the $U(1)_{em}'$
symmetry can be spontaneously broken to give the dark photon a
mass~\cite{Baumgart:2009tn}. The familiar $U(1)_{em}$ is, of course,
kept unbroken. We choose the mass of the mirror photon to be
${\mathcal O}(100)$ MeV.

The massive dark photon is coupled to the SM fermions through the
kinetic mixing $\varepsilon_\gamma/2 F^{\mu\nu}F_{\mu\nu}'$ and
therefore decays to $e^\pm $ pair. The decay rate is given by
$\Gamma_{\gamma'}=\alpha_{em} \varepsilon_\gamma^2 m_{\gamma'}/3$.
The lifetime of the dark photon is
\begin{eqnarray}
\tau_{\gamma'}\approx\left(\frac{50\rm{MeV}}{m_{\gamma'}}\right)\left
(\frac{7\times 10^{-11}}{\varepsilon_\gamma}\right)^2 \rm{sec} \ .
\end{eqnarray}
For $m_{\gamma'}=50$ MeV, $\varepsilon_\gamma > 7 \times 10^{-11}$ is
needed to make the dark photon lifetime be shorter than $1$ sec.

QED precision measurements provides constraints on the coupling $\varepsilon_\gamma$.
 The most important constraint comes from the measurement
of the muon magnetic moment, which gives an upper bound
$\varepsilon_\gamma^2< 2\times 10^{-5} (m_{\gamma'}/100 \mbox{MeV})^2$
~\cite{Pospelov:2008zw} .

The mass of the mirror photon is induced by the nonvanishing VEVs
of charged components of the mirror doublet Higgs bosons.  The
other non-trivial consequence of breaking dark $U(1)_{em}'$ is the
charged mirror particles will mix with neutral particles since
$\langle H'^\pm\rangle\neq0$. These small mixings allow the mirror
proton to radiatively decay to mirror neutron, i.e., $p^\prime \to
n^\prime \gamma^\prime$ and the mirror electron to decay as
$e^\prime\to \gamma^\prime \nu$. The size of this mixing can
roughly be estimated as
$U_{p^\prime n^\prime}'\simeq \frac{v^\prime_{\pm}y_d}{v^\prime_{wk}y_u} \approx
\frac{100{\rm MeV}y_d}{100 {\rm TeV}y_u}\approx 10^{-6}\cdot 2\tan \beta $ for
the case we are interested.
For tan$\beta \sim 50$, this gives $U_{p^\prime n^\prime}'\sim 10^{-4}$.

The decay rate for the process $p^\prime\to n^\prime \gamma^\prime$ is given by
\begin{equation}
 \Gamma\approx \frac{\alpha_{em} U^{\prime2}_{p^\prime n^\prime} m_{p^\prime}}{4}F(\frac{m_{n^\prime}}{m_{p^\prime}},
\frac{m_{\gamma^\prime}}{m_{p^\prime}})\,,
\end{equation}
where $F(x,y)=\frac{(1-x^2+y^2)(1-x^2-y^2)}{y^2}\left[(1-(x+y)^2)(1-(x-y)^2)\right]^{1/2}$.
For the case $\tan\beta'=200$, one can obtain the lifetime of mirror proton $\tau_{p^\prime}\approx 10^{-15}$ sec.
For the mirror electron decay, $e^\prime\to \gamma^\prime \nu$, inputing
$m_{e^\prime}=0.13 $ GeV and $U_{e^\prime \nu}\approx {v^\prime_{\pm}}/{v^\prime_{wk}}U_{\nu^\prime \nu}  \approx 10^{-9}$ which agrees with the numerical result, its lifetime is estimated to be $5\times 10^{-2}$ sec.
Both of them decay before the epoch of BBN.

\subsection{Constraints from neutrinoless double beta decay}
The dark neutrino contributes to the $0\nu\beta\beta$ decay process by exchange $\nu'$
 due to the mixing with the active neutrino $\nu_e$. Conventionally one parameterizes the experimental bounds on
 $0\nu\beta\beta$ decay as a limit on the ``effective" neutrino mass $m^{eff}_\nu$. For light neutrinos, $m^{eff}_\nu$
is defined as $\Sigma_i U_{ei}^2 m_i$. The current upper bound is $m^{eff}_\nu< 0.5$ eV from the
$^{76}$Ge $0\nu\beta\beta$ experiment~\cite{doublebeta}.

When the dark sterile neutrino masses are heavy compared to $\mathcal O(100)$ MeV, like the case we are considering here,
one gets the contribution to the effective mass~\cite{Bamert:1994qh}
\begin{equation}
m^{eff}_\nu=\frac{U^2_{\nu\nu'}{\bar q_F}^2}{3m_{\nu'}}\,,
\end{equation}
where ${\bar q_F}$ is the nucleon Fermi momentum
and its value is taken as $\bar q\approx 60$
MeV~\cite{Bamert:1994qh}. So, inputing $U_{\nu\nu'}=10^{-3},
m_{\nu'}=136$ MeV, one finds that this parameter choice has tension
to satisfy the experimental limit. However, if there is cancellation between the
contributions from $\nu'$'s and light neutrinos, the constraint
could be relaxed. Furthermore, considering the structure of the total neutrino mass
matrix in Eq.~(\ref{inv}), we notice that the light SM neutrino
mass matrix is $M_\nu= \mu +\mu' (v_{\rm wk}/v'_{wk})^2$ due to
the exact mirror symmetric Yukawa couplings. This means there is
no family-mixing between the light neutrinos and the sterile dark
neutrinos, $U_{ij'}\propto \delta_{ij'}$ as noted already. The
mirror neutrino contribution to $0\nu\beta\beta$ decay  is
proportional to the $e'e'$ component of the sterile dark neutrino
mass matrix $(M^\prime_D M_R^{-1} M^{\prime T}_D)_{e'e'}$. The
active neutrino mass matrix, on the other hand, is given by the
matrix $\mu^\prime$. Therefore they are unrelated and we are free to
choose the $e'e'$ component of the mirror neutrino mass matrix
without affecting the active neutrino mixings. Choosing  a tiny
or even vanishing value for this element can guarantee the model
to avoid the experimental limits.

\subsection{Constraints of self interaction cross-section of dark
matter} In our model since mirror neutron is the dark matter, it
will have strong scattering against other mirror neutrons. There
are two sources for this scattering to arise from: (i) strong
scattering and (ii) electromagnetic scattering. As regards strong
scattering, we note that familiar low energy neutron scattering of neutrons
off protons has a cross-section of order $\sigma_{np}\sim
10^{-24}$ cm$^2$ and by isospin symmetry, the $\sigma_{nn}\simeq
\sigma_{np}$. Since in the mirror sector, $\Lambda'_{QCD}\sim 10
\Lambda_{QCD}$, we expect the cross section $\sigma_{n'n'}\sim
10^{-26}$ cm$^2$. Note that the upper bound on the dark matter
self interaction cross section is given by~\cite{SDM}
$\sigma/m_{dark matter} < 1.25$ cm$^2$ gram$^{-1}$ which can
interpreted as $\sigma_{n'n'}\leq 10^{-23}$ cm$^2$ in our case
where the dark matter mass is about 5 GeV. Coming to the electromagnetic
scattering cross-section, we note that $n'$ has no mirror electric
charge but a mirror magnetic moment given roughly by $\mu_{n'}\sim
e\frac{k}{m_{n'}}$, where k can be estimated as $m_{dark
matter}\times v$, and the velocity $v/c$ is roughly $10^{-3}$.
Therefore the electromagnetic vertex which goes into self
scattering has a strength of order $10^{-3}e$. Now, since the
momentum transfer is much smaller than the mass of the dark
photon, the vertex of the four-dark matter interaction can be
written as $10^{-6}e^2/m^2_{\gamma'}$. The cross section is
estimated to be $(10^{-6}e^2/m^2_{\gamma})^2\times m^{2}_{n'}/(4\pi)$ which is at most $10^{-35}$ cm$^2$, which is much
below the bullet cluster upper bound.

\section{Direct detection}
Asymmetric dark matter can be detected directly by observing the
nucleus recoil at low background experiments. The effective
operators for DM-nucleon interaction can be generalized as
\begin{equation}
\overline\chi \Gamma_1\chi \overline N\Gamma_2N \,,
\end{equation}
where $\Gamma_{1,2}$ can be the combinations of scalar,
pseudoscalar, vector, axial, tensor or
pseudo-tensor~\cite{Kurylov:2003ra}. In the non-relativistic limit,
all the interactions can be reduced to two terms, spin-independent
(SI) and spin-dependent (SD).

In our model, the direct detection process can be induced by the
interaction between the Higgs bosons in the two sectors and
kinetic mixing between the gauge bosons.

First, let's consider the direct detection via the Higgs
interaction $f|H|^2|H'|^2$. After the spontaneous symmetry
breaking in the two sectors, this term generates an effective
four-fermion interaction between nucleons and mirror nucleons,
which can be written as
\begin{equation}
{\cal L}_{eff} = \delta_f \overline n' n' \overline N N\ ,
\end{equation}
where $\delta_f\simeq({f m_N m_{n'}})/({10m^2_{h}
m^2_{h^\prime}})$~\cite{Shifman:1978zn}. The total cross section of
the elastic scattering between $n'$ and the nucleon can be written
as
\begin{equation}
\sigma_{Nn'}\approx \frac{\mu_r^2}{\pi} \delta_f^2 \approx 10^{-29}
f^2 {\rm ~GeV}^{-2} \approx 10^{-57} f^2 {\rm ~cm}^{2}\ ,
\end{equation}
where $\mu_r=m_N m_{n'} /(m_N+m_{n'})$ is the reduced mass of
$n'$ and nucleon $N$, $m_h$ is the mass of a Higgs boson from the SM
sector which is set to be 100 GeV and $m_{h'}$ from the mirror sector
is set to be 100 TeV. This is well below
the present upper bound of $10^{-39}$ cm$^2$ for around 5 GeV dark matter.

The kinetic mixing between photon and mirror photon plays an important role
in the direct detection. The local velocity of dark matter is assumed to be
$\sim 200$ km/s, so for a 5 GeV dark matter its kinetic energy is about 1
KeV, which is much smaller than the mass of the mirror photon which
is assumed to be 50 MeV. Therefore, the interaction between the
nucleon and the mirror neutron can also be viewed as a point-like
interaction. The mirror neutron interacts with the mirror photon
through its anomalous magnetic moment and also through its mixing
with the mirror proton due to the breaking of the mirror
electromagnetic $U(1)$ symmetry.

Since $u'$ and $d'$ are heavy, one can use constituent quark model
to estimate the magnetic moment of the mirror neutron.
\begin{equation}
\mu_n'=\frac{4}{3}\mu_d'-\frac{1}{3}\mu_u'\ ,
\end{equation}
where $\mu'_{(u,d)}=eQ_{u,d}/2m_{u',d'}$ are magnetic moments of $u'$
and $d'$ where $Q_u=2/3$ and $Q_d=-1/3$ are the mirror charges of
$u'$ and $d'$. In the nonrelativistic regime the total cross section
between the nucleon and the mirror neutron induced by this magnetic
interaction can be estimated as
\begin{equation}
\sigma^{mag} \approx \frac{\alpha_{em}^2 \pi\varepsilon_\gamma^2|\vec
q|^4}{(m_{n'}+m_N)^2 {m_{\gamma'}}^4}\lesssim
4\times 10^{-8}\varepsilon_\gamma^2~ {\rm
GeV}^{-2}\lesssim 10^{-35}\varepsilon_\gamma^2~{\rm cm}^2\ ,
\end{equation}
where $\vec q$ is the momentum transfer during the collision which
is about 5 MeV. 
A less important contribution comes from the mixing between
the $n'$ and $p'$. The cross section induced by this mixing can be written as
\begin{equation}
\sigma^{n^\prime p^\prime} \approx
\frac{\mu_r^2}{\pi}\left(\frac{\varepsilon_\gamma e^2 {U^{\prime 2
}_{n^\prime p^\prime}}}{m_{\gamma'}^2}\right)^2\approx
10^{-14}\varepsilon_\gamma^2 ~{\rm
GeV}^{-2}\approx10^{-41}\varepsilon^2_\gamma ~{\rm cm}^2 \ .
\end{equation}
Recalling the upper bound on $\epsilon_\gamma$ from QED
precision measurements given above, we conclude that the cross section is well below the upper bound set by
direct detection experiments and our dark
matter could be accessible to direct search experiments in future.

From Eq.~(\ref{mix}) in Appendix, one can see that interactions through $Z$ and
$Z'$ between the two sectors are either suppressed by the nature of
the magnetic interaction or by the $M_Z^2/M_{Z'}^2$ and then are negligible.

\section{Conclusions}
In this work, we have proposed that the dark matter of the
Universe be identified with the lightest baryon of a possible
mirror duplicate of the standard model with the only difference
between the two sectors being in the symmetry breaking patterns.
Prior to spontaneous symmetry breaking, this model has no free
parameters due to mirror symmetry. The lightest dark nucleon is
stable due to the mirror analog of baryon number and becomes the
dark matter. It is an asymmetric dark matter with its anti-mirror
baryon part suppressed in a manner analogous to the
matter-anti-matter asymmetry in the standard model sector. The
introduction of a common set of right-handed neutrinos connecting
the two sectors allows a common mechanism for the genesis of the
matter-anti-matter asymmetry in both sectors thereby helping us to
understand why the dark matter and normal baryon contribution to
the energy density of the Universe are not too different from each
other. One only has to make the assumption that the dark nucleon
is five times heavier than the familiar nucleon -- an assumption
that is easily understood if the electroweak scales in the two
sectors are different. We show that this model can be consistent
with the constraints of BBN and neutrinoless double beta decay.
The mirror photon in our model is massive but mixes with the
normal photon to avoid the BBN constraints.

\section*{Acknowledgments}
This work was partially supported by the U. S. Department of Energy via grant DE-FG02-
93ER-40762. The works of RNM is supported by the NSF grant PHY-0652363.
We like to thank S. Blanchet, Thomas Cohen, Xiangdong Ji, S. Nussinov, Alexei Yu Smirnov for discussions.

\renewcommand{\theequation}{A\arabic{equation}}
\setcounter{equation}{0} 

\section*{Appendix A: Kinetic mixing between gauge bosons}

The kinetic mixing between the $U(1)$ gauge bosons $B$ and $B'$
will induce kinetic mixings of $\gamma$ and $Z$ with their mirror
partners due to the spontaneous symmetry breaking in both sectors.
Therefore the Lagrangian for kinetic mixing can be written as
\begin{equation}
{\cal L}_{kin}^{mix} = \frac{\varepsilon_B}{2} [\cos^2\theta_W
F^{\mu\nu}F'_{\mu\nu}-\sin\theta_W\cos\theta_W(
F^{\mu\nu}Z'_{\mu\nu}+ Z^{\mu\nu}F'_{\mu\nu})+\sin^2\theta_W
Z^{\mu\nu}Z'_{\mu\nu}]\ ,
\end{equation}
where $\theta_W$ is the Weinberg angle. Since exact mirror symmetry
is assumed the two sectors share the same Weinberg angle at tree
level. Using the fact that $M^2_{Z'}\gg M_Z^2\gg m_{\gamma'}^2$, to
the leading order of $\epsilon_B$ the following redefinitions of
$A_\mu$, $A_\mu'$, $Z_\mu$ and $Z_\mu'$ diagonalize both the kinetic
terms and the mass terms of gauge bosons:
\begin{eqnarray}\label{mix}
A_\mu &\longrightarrow& A_\mu + \varepsilon_\gamma A'_\mu -
\sqrt{\varepsilon_\gamma \varepsilon_Z}Z'_\mu\ ; \nonumber\\
Z_\mu &\longrightarrow& Z_\mu + \varepsilon_Z Z_\mu' +
\sqrt{\varepsilon_\gamma\varepsilon_Z}\frac{M_{A'}^2}{M_Z^2}A_\mu'\
;\nonumber\\
A'_\mu &\longrightarrow& A'_\mu -
\sqrt{\varepsilon_\gamma\varepsilon_Z}Z_\mu\ ;\nonumber\\
Z'_\mu &\longrightarrow& Z'_\mu -
\varepsilon_Z\frac{M_Z^2}{M_{Z'}^2} Z_\mu\ ,
\end{eqnarray}
where $\varepsilon_\gamma = \varepsilon_B \cos^2\theta_W$ and
$\varepsilon_{Z} = \varepsilon_B \sin^2\theta_W$.


\begin{thebibliography}{99}
\bibitem{Sakharov:1967dj}
  A.~D.~Sakharov,
  Pisma Zh.\ Eksp.\ Teor.\ Fiz.\  {\bf 5}, 32 (1967)
  [JETP Lett.\  {\bf 5}, 24 (1967\ SOPUA,34,392-393.1991\ UFNAA,161,61-64.1991)].

\bibitem{Fukugita:1986hr}
  M.~Fukugita and T.~Yanagida,
  Phys.\ Lett.\  B {\bf 174}, 45 (1986).


\bibitem{seesaw}
P.~Minkowski, Phys.\ Lett.\ B~{\bf 67} (1977) 421; M. Gell-Mann, P.  Ramond and R. Slansky, in  {\em Supergravity},
eds.~D.Z.  Freedman  and  P.~van Nieuwenhuizen  (North-Holland,  Amsterdam,
1979); T. Yanagida, in  Proc.\ of  the {\em  Workshop on  the Unified
Theory and the  Baryon Number in the Universe},  Tsukuba, Japan, 1979,
eds.\ O.~Sawada and  A.~Sugamoto;
S. L. Glashow, {\em The future of elementary particle physics}, in {\em Proceedings
of the 1979 Carg`ese Summer Institute on Quarks and Leptons}, Plenum Press, New York, 1980;
R.~N.~Mohapatra and G.~Senjanovi\'c, Phys.\ Rev.\ Lett.\ {\bf 44},912 (1980).



\bibitem{Kuzmin:1985mm}
  V.~A.~Kuzmin, V.~A.~Rubakov and M.~E.~Shaposhnikov,
  Phys.\ Lett.\  B {\bf 155}, 36 (1985).



\bibitem{mirror} T. D. Lee and C. N. Yang, Phys. Rev. {\bf 104}, 254
(1956); K. Nishijima, private communication; Y. Kobzarev, L. Okun
and I. Ya Pomeranchuk, Yad. Fiz. {\bf 3}, 1154 (1966);  M. Pavsic,
Int. J. T. P. {\bf 9}, 229 (1974); S. I. Blinnikov and M. Y.
Khlopov, Astro. Zh. {\bf 60}, 632 (1983); R. Foot, H. Lew and R.
R. Volkas, Phys. Lett. {\bf B272}, 67 (1991); Mod. Phys. Lett. {\bf A7}, 2567 (1992);
R. Foot and R. Volkas,
Phys. Rev. {\bf D 52}, 6595 (1995); Z. Berezhiani and R. N.
Mohapatra, Phys. Rev. {\bf D 52}, 6607 (1995); Z. Berezhiani, A.
Dolgov and R. N. Mohapatra, Phys. Lett. {\bf B 375}, 26 (1996);
Z. Silagadze, Phys. At. Nucl. {\bf 60}, 272 (1997); for a review of the literature, see
L.~B.~Okun,
  Phys.\ Usp.\  {\bf 50}, 380 (2007)
  [arXiv:hep-ph/0606202].


  \bibitem{Buchmuller:2004nz}
W.~Buchmuller, P.~Di Bari and M.~Plumacher,
  Annals Phys.\  {\bf 315}, 305 (2005)
  [arXiv:hep-ph/0401240].

\bibitem{Covi:1996wh}
  L.~Covi, E.~Roulet and F.~Vissani,
  Phys.\ Lett.\  B {\bf 384}, 169 (1996)
  [arXiv:hep-ph/9605319];
M.~Flanz, E.~A.~Paschos, U.~Sarkar and J.~Weiss,
  Phys.\ Lett.\  B {\bf 389}, 693 (1996)
  [arXiv:hep-ph/9607310];
    A.~Pilaftsis and T.~E.~J.~Underwood,
  Nucl.\ Phys.\  B {\bf 692}, 303 (2004)
  [arXiv:hep-ph/0309342];
  A.~Pilaftsis and T.~E.~J.~Underwood,
  Phys.\ Rev.\  D {\bf 72}, 113001 (2005)
  [arXiv:hep-ph/0506107].



\bibitem{adm} For other ideas on connecting the baryonic matter density to dark matter density, see
  S.~Nussinov,
  Phys.\ Lett.\  B {\bf 165}, 55 (1985);
  S.~M.~Barr,
  Phys.\ Rev.\  D {\bf 44}, 3062 (1991);
  D.~B.~Kaplan,
  Phys.\ Rev.\ Lett.\  {\bf 68}, 741 (1992);
  S.~M.~Barr, R.~S.~Chivukula and E.~Farhi,
  Phys.\ Lett.\  B {\bf 241}, 387 (1990);
  S.~Dodelson, B.~R.~Greene and L.~M.~Widrow,
  Nucl.\ Phys.\  B {\bf 372}, 467 (1992);
  V.~A.~Kuzmin,
  Phys.\ Part.\ Nucl.\  {\bf 29}, 257 (1998)
  [Fiz.\ Elem.\ Chast.\ Atom.\ Yadra {\bf 29}, 637 (1998\ PANUE,61,1107-1116.1998)]
  [arXiv:hep-ph/9701269];
  R.~N.~Mohapatra, S.~Nussinov and V.~L.~Teplitz,
  Phys.\ Rev.\  D {\bf 66}, 063002 (2002)
  [arXiv:hep-ph/0111381].
  M.~Fujii and T.~Yanagida,
  Phys.\ Lett.\  B {\bf 542}, 80 (2002)
  [arXiv:hep-ph/0206066];
  R.~Foot,
  Int.\ J.\ Mod.\ Phys.\  D {\bf 13}, 2161 (2004)
  [arXiv:astro-ph/0407623].
  D.~Hooper, J.~March-Russell and S.~M.~West,
  Phys.\ Lett.\  B {\bf 605}, 228 (2005)
  [arXiv:hep-ph/0410114];
  G.~R.~Farrar and G.~Zaharijas,
  Phys.\ Rev.\ Lett.\  {\bf 96}, 041302 (2006)
  [arXiv:hep-ph/0510079];
  S.~B.~Gudnason, C.~Kouvaris and F.~Sannino,
  Phys.\ Rev.\  D {\bf 73}, 115003 (2006)
  [arXiv:hep-ph/0603014];
  R.~Kitano, H.~Murayama and M.~Ratz,
  Phys.\ Lett.\  B {\bf 669}, 145 (2008)
  [arXiv:0807.4313 [hep-ph]];
  L.~Roszkowski and O.~Seto,
  Phys.\ Rev.\ Lett.\  {\bf 98}, 161304 (2007)
  [arXiv:hep-ph/0608013]l
  O.~Seto and M.~Yamaguchi,
  Phys.\ Rev.\  D {\bf 75}, 123506 (2007)
  [arXiv:0704.0510 [hep-ph]];
 N.~Sahu and U.~Sarkar,
 Phys. Rev.  D {\bf 78}, 115013 (2008)
 [arXiv:0804.2072 [hep-ph]];
 K.~Kohri, A.~Mazumdar and N.~Sahu,
 Phys. Rev.  D {\bf 80}, 103504 (2009)
 [arXiv:0905.1625 [hep-ph]];
   P.~H.~Gu, U.~Sarkar and X.~Zhang,
  Phys.\ Rev.\  D {\bf 80}, 076003 (2009)
  [arXiv:0906.3103 [hep-ph]].
 K.~Kohri, A.~Mazumdar, N.~Sahu and P.~Stephens,
 Phys. Rev.  D {\bf 80} (2009) 061302
 [arXiv:0907.0622 [hep-ph]];
 P.~H.~Gu and U.~Sarkar,
  arXiv:0909.5463 [hep-ph].
 D.~E.~Kaplan, M.~A.~Luty and K.~M.~Zurek,
  Phys.\ Rev.\  D {\bf 79}, 115016 (2009)
  [arXiv:0901.4117 [hep-ph]].
  D.~E.~Kaplan, G.~Z.~Krnjaic, K.~R.~Rehermann and C.~M.~Wells,
  arXiv:0909.0753 [hep-ph].
  G.~D.~Kribs, T.~S.~Roy, J.~Terning and K.~M.~Zurek,
  arXiv:0909.2034 [hep-ph].
  T.~Cohen and K.~M.~Zurek,
  arXiv:0909.2035 [hep-ph].


\bibitem{mdm} See for instance,   H.~M.~Hodges,
  Phys.\ Rev.\  D {\bf 47}, 456 (1993); R.~N.~Mohapatra and V.~L.~Teplitz,
  Phys.\ Lett.\  B {\bf 462}, 302 (1999);
  Phys.\ Rev.\  D {\bf 62}, 063506 (2000);
 Z.~Berezhiani, D.~Comelli and F.~L.~Villante,
  Phys.\ Lett.\  B {\bf 503}, 362 (2001);  R.~N.~Mohapatra, S.~Nussinov and V.~L.~Teplitz,
  Phys.\ Rev.\  D {\bf 66}, 063002 (2002);  A.~Y.~Ignatiev and R.~R.~Volkas,
  Phys.\ Rev.\  D {\bf 68}, 023518 (2003);  Z.~Berezhiani, P.~Ciarcelluti, D.~Comelli and F.~L.~Villante,
  Int.\ J.\ Mod.\ Phys.\  D {\bf 14}, 107 (2005); P.~Ciarcelluti,
  Int.\ J.\ Mod.\ Phys.\  D {\bf 14}, 187 (2005).

\bibitem{zurab}
L.~Bento and Z.~Berezhiani,
  Phys.\ Rev.\ Lett.\  {\bf 87}, 231304 (2001)


\bibitem {inverse} R. N. Mohapatra, Phys. Rev. Lett. {\bf 56}, 561 (1986);
R. N. Mohapatra and J. W. F. Valle, Phys. Rev. D {\bf 34}, 1642 (1986).


\bibitem{tripletlepto}
  T.~Hambye and G.~Senjanovic,
  Phys.\ Lett.\  B {\bf 582}, 73 (2004)
  [arXiv:hep-ph/0307237];
  S.~Antusch and S.~F.~King,
  Phys.\ Lett.\  B {\bf 597}, 199 (2004)
  [arXiv:hep-ph/0405093].




\bibitem{Cyburt:2004yc}
  R.~H.~Cyburt, B.~D.~Fields, K.~A.~Olive and E.~Skillman,
  Astropart.\ Phys.\  {\bf 23}, 313 (2005)
  [arXiv:astro-ph/0408033];
  C.~Amsler {\it et al.}  [Particle Data Group],
  Phys.\ Lett.\  B {\bf 667}, 1 (2008).

  \bibitem{bere} Z. Berezhiani, A.
Dolgov and R. N. Mohapatra, Phys. Lett. {\bf B 375}, 26 (1996).

\bibitem{Baumgart:2009tn}
  M.~Baumgart, C.~Cheung, J.~T.~Ruderman, L.~T.~Wang and I.~Yavin,
  JHEP {\bf 0904}, 014 (2009)
  [arXiv:0901.0283 [hep-ph]].

\bibitem{Pospelov:2008zw}
  M.~Pospelov,
  Phys.\ Rev.\  D {\bf 80}, 095002 (2009)
  [arXiv:0811.1030 [hep-ph]].

\bibitem{doublebeta}
H. V. Klapdor-Kleingrothaus {\it et al.}, Eur. Phys. J. A 12, 147 (2001); C. E. Aalseth {\it et
al.}, Phys. Rev. {\bf D 65}, 092007 (2002).

\bibitem{Bamert:1994qh}
  P.~Bamert, C.~P.~Burgess and R.~N.~Mohapatra,
  Nucl.\ Phys.\  B {\bf 438}, 3 (1995)
  [arXiv:hep-ph/9408367]; P.~Benes, A.~Faessler, F.~Simkovic and S.~Kovalenko,
  Phys.\ Rev.\  D {\bf 71}, 077901 (2005)
  [arXiv:hep-ph/0501295].

\bibitem{SDM}   S.~W.~Randall, M.~Markevitch, D.~Clowe, A.~H.~Gonzalez and M.~Bradac,
  arXiv:0704.0261 [astro-ph].



\bibitem{Kurylov:2003ra}
  A.~Kurylov and M.~Kamionkowski,
  Phys.\ Rev.\  D {\bf 69}, 063503 (2004)
  [arXiv:hep-ph/0307185].

\bibitem{Shifman:1978zn}
  M.~A.~Shifman, A.~I.~Vainshtein and V.~I.~Zakharov,
  Phys.\ Lett.\  B {\bf 78}, 443 (1978).


\end{thebibliography}
\end{document}